\documentclass[useAMS,usegraphicx]{mn2e}

\title[$N_H$ short time variability in UGC 4203]{{\em Chandra} monitoring of UGC~4203: the structure of the X-ray absorber}
\author[G. Risaliti et al.]
{G. Risaliti,$^{1,2}$ 
M.~Elvis,$^2$ 
S. Bianchi$^3$, 
G. Matt$^3$\\
$^1$ INAF - Osservatorio Astrofisico di Arcetri, L.go E. Fermi 5,
Firenze, Italy\\
$^2$ Harvard-Smithsonian Center for Astrophysics, 60 Garden St. 
Cambridge, MA 02138 USA {E-mail: grisaliti@cfa.harvard.edu}\\
$^3$ Dipartimento di Fisica, Universit\`a degli Studi ``Roma Tre'',
Via della Vasca Navale 84, I-00146 Roma, Italy\\
}
\begin{document}

\date{Released Xxxx Xxxxx XX}

\pagerange{\pageref{firstpage}--\pageref{lastpage}} \pubyear{2002}

\maketitle

\label{firstpage}

\begin{abstract}
We present a {\em Chandra} monitoring campaign of the highly variable 
Seyfert galaxy UGC~4203 (the ``Phoenix Galaxy'') which revealed variations
in the X-ray absorbing column density on time scales of two weeks.
This is the third, clear case, after NGC~1365 and NGC~7582, of dramatic 
N$_H$ variability on short time scales observed in 
a ``changing~look'' source, i.e. an AGN observed in the past in both a
reflection-dominated and a Compton-thin state.
The inferred limits on the distance of the X-ray absorber from the center
suggest that the X-ray ``torus'' could be one and the same with the broad emission line region.
This scenario, first proposed for an ``ad-hoc'' picture for NGC~1365, may be the 
common structure of the circumnuclear medium in AGN. 

\end{abstract}


\section{Introduction}
The structure, size and composition of the circumnuclear medium of Active Galactic Nuclei (AGN) 
is the subject of several studies at different wavelengths. In particular, X-ray absorption 
variability has proven to be a powerful tool in order to establish the physical properties
of the X-ray absorber/reflector: the almost ubiquitous absorption variability detected 
in bright absorbed AGN (Risaliti et al.~2002) is in itself an indication of a clumpy structure of the absorber. The time scale of such variations provide constraints on the dimensions and the 
distance from
the central black hole of the obscuring clouds, under the assumption of Keplerian motion.
The best characterization so far of the physical parameters of the X-ray absorber has been possible for the AGN in NGC~1365. This source has been monitored several times in the last few years
with {\em Chandra}, {\em XMM-Newton}, {\em Suzaku}, and showed extreme variability in the X-rays:
changes from Compton-thick to Compton-thin states on time scales from weeks (Risaliti et al.~2005)
to days (Risaliti et al.~2007) to $\sim10$~hours (Risaliti et al.~2009A) and eclipses due
to Compton-thin clouds ($N_H\sim$a few $10^{23}$~cm$^{-2}$) on similar time scales (Risaliti et al.~2009B).
These observations strongly support the view of an X-ray absorber made of clouds with $N_H$
from $10^{23}$ to $>$10$^{24}$~cm$^{-2}$, densities 10$^{10}$-10$^{11}$~cm$^{-3}$ and distance from the center 
of hundreds of gravitational radii. These are all values typical of the broad emission line clouds (BELC). 
The global picture emerging for NGC~1365 is therefore quite different from the standard view of
the AGN unified models (Antonucci~1993):
here, the X-ray absorber seems not to be associated to the parsec-scale dusty ``torus'' (Krolik \& Begelman~1988) obscuring the BELC, but to the BELC themselves.

This scenario appears quite well justified for the single case of NGC~1365, whose 
extreme variability is so far unique among AGN. Two explanations of this uniqueness are possible:
(1) NGC~1365 is intrinsically different from (most of) the other AGN; (2) NGC~1365 is an ``average'' AGN with favourable observational conditions (e.g. a particular line of sight through
a torus made of clouds of different $N_H$, as in Risaliti et al.~2005).

In order to investigate this issue, the obvious strategy is to look for similar  variations
in other obscured AGN. 

The most extreme known cases of spectral variations in AGN are those of the ``changing~look''
AGN (Matt et al.~2003), i.e. sources with past X-ray observations in both a Compton-thin and a 
reflection-dominated state. Members of this class, besides NGC~1365, are NGC~6300 (Guainazzi.~2002), UGC~4203 (Guainazzi et al.~2002), NGC~2992 (Gilli et al.~2000), NGC~7674 (Bianchi et al.~2005), NGC~7582 (Piconcelli et al.~2007). In the latter
case, a recent {\em Suzaku} campaign revealed N$_H$ variations in time scales of a day (Bianchi et al.~2009), 
making NGC~7582 a second case of
a ‘‘changing look'' AGN with fast N$_H$ variability. 

The main open question about these sources is whether the observed dramatic spectral changes
are
due to variability of the intrinsic emission or of the circumnuclear absorber. The case of
NGC~2992 is clear: multiple observations on a $\sim20$~years period strongly suggest 
an intrinsic fading of the X-ray source (Gilli et al.~2000). We note that this is the only 
case among the ``changing look'' sources with a column density in the Compton-thin state
of the order of 10$^{22}$~cm$^{-2}$, while the other sources have $N_H$$>$$10^{23}$~cm$^{-2}$.

Here we present the results of a {\em Chandra} monitoring campaign
of one of the three remaining sources, UGC~4203 (other names: MKN~1210 and ``Phoenix Galaxy'').
Another {\em Chandra} campaign on one of the last two sources, NGC~6300, will be discussed in a future paper.

\section{Past observations}
UGC~4203 has been observed several times in the hard X-rays.
The two observations which led to its classification as a ``changing look'' AGN, reported in 
Guainazzi et al.~2002, were performed with ASCA in 1998 and {\em XMM-Newton} in 2001.
A further observation, has been performed simultaneously 
with {\em XMM-Newton} by {\em BeppoSAX}. 
More recently, the source has been observed with {\em Suzaku} in 2007, and
with {\em Swift}-XRT for four times between 2006 and 2007. 
In order to obtain a complete view of the variability history of UGC~4203,
we analyzed the {\em Swift-XRT} short observations (about two ks each) in order to
estimate the flux and, if possible, the basic spectral parameters. We followed
the standard procedure suggested by the {\em Swift} Data Center for the reduction and
calibration. In all cases the net 2-10~keV spectral counts are below 100,
allowing only a basic spectral analysis. 

A summary of the spectral parameters obtained
in these observations, assuming an absorbed power~law model, is shown in Table~1.
\begin{table}
\caption{Spectral parameters from past X-ray observations}
\centerline{\begin{tabular}{lcccccc}
Instr. & Date & Counts & F$_X^a$ & $\Gamma^b$ & $N_H^c$ \\
\hline
ASCA   & 18/11/95 &320        & 0.2$^e$    & 2.0$^d$             &  --                  \\
XMM    & 05/05/01 &$>$10$^4$  & 2.6        & 2.0$^{+0.2}_{-0.2}$ & 2.1$^{+0.2}_{-0.2}$ \\
Suzaku & 27/05/07 &$>$10$^4$  & 1.4        & 1.9$\pm0.2$&          3.3$^{+0.2}_{-0.2}$  \\
Swift 1& 06/10/06 &88         & 1.6$\pm0.4$& 1.9$^d$&              1.7$^{+0.5}_{-0.4}$  \\
Swift 2& 22/01/07 &32         & 1.2$\pm0.5$& 1.9$^d$&              2.5$^{+1.5}_{-1.0}$  \\
Swift 3& 25/02/07 &43         & 0.9$\pm0.5$& 1.9$^d$&              1.9$^{+1.0}_{-0.6}$  \\
Swift 4& 29/04/08 &35         & 2.3$\pm0.8$& 1.9$^d$&              3.4$^{+1.2}_{-1.6}$  \\
\hline
\end{tabular}}
Results form past hard X-ray observations of UGC~4203. 
$^a$: 2-10~keV unabsorbed flux, in units of 10$^{-11}$~erg~s$^{-1}$~cm$^{-2}$; 
$^b$: 2-10~keV photon index; $^c$: absorbing column density, in units of 10$^{23}$~cm$^{-2}$;
$^d$: fixed. $^e$: flux of the observed reflection-dominated spectrum.
References: ASCA and XMM: Guainazzi et al.~2002; {\em Swift}: this work; {\em Suzaku}:
 Matt et al.~2009.
\end{table}
\section{Chandra observations: reduction and analysis}
The five 10~ks
observations were performed with the ACIS-S instrument on-board {\em Chandra} between
January~15 and February~06, 2008. The first
three observations are almost consecutive, the fourth one is about two days after, and
the fifth 17 days after the fourth. 

We reduced and analyzed the data using the CIAO~4.2 package, and the most updated
calibrations, following the standard procedure developed by the {\em Chandra} Data 
Center\footnote{http://cxc.harvard.edu/ciao/index.html}.
The background was extracted in an empty region in the ACIS-S field of view. The data
were checked for possible alterations due to pile-up. Despite the high unabsorbed flux
of the source (of the order of 10$^{-11}$~erg~s$^{-1}$~cm$^{-2}$), no significant pile-up is
expected, since the soft ($<3$~keV) 
photons from the AGN are absorbed by a column density
of the order of 2-3$\times$10$^{23}$~cm$^{-2}$.
The spectral analysis has been performed with the Xspec~12 package (Arnaud et al.~1996).

A visual inspection of the five spectra (Fig.~1A,~B) reveals  a soft component at E$<$2~keV 
constant within a few
per cent, and a second, hard component at E$>$2~keV with clear, strong variations
between the fifth observation and the others. 

We fitted the five spectra separately with a model consisting of the following 
components:\\
1. a soft spectrum produced by collisionally-ionized diffuse gas, with two
different temperatures and densities (modeled through
the APEC model\footnote{http://cxc.harvard.edu/atomdb/} in XSPEC), 
and a power law, representing a possible scattering component from
the central AGN. Our aim is to obtain a statistically acceptable representation of the
soft emission, in order to fix it in the analysis of the variability of the hard component.
An alternative, simpler approach would be an analysis of the spectrum at energies E$>$3~keV,
neglecting the soft part of the spectrum. This would however prevent us from rejecting
models reproducing the hard spectrum, but whose extrapolation at low energies over-predicts
the observed counts. The fitting analysis shows that all the three components
are needed to obtain a good representation of the observed data. The thermal 
component has a temperature kT=0.7$\pm0.2$~keV and the power law slope and
normalizations are $\Gamma_S=2.05\pm0.15$ and $N_S=2.0_{-0.3}^{+0.4}$$\times$10$^{-5}$~keV~s$^{-1}$~cm$^{-2}$~keV$^{-1}$ 
(corresponding to a flux of a few per cent of that of the
intrinsic AGN continuum, see below). These are all typical parameters for the soft
emission of local AGN. The metallicities of the photoionized component span a large range
(from 0.5 to 20 solar) and are poorly constrained. This is analogous to what 
is obtained in global fits of the soft diffuse emission of other AGN. A spatially
resolved analysis of these cases, when possible (i.e. in lower-redshift sources) show
that the total soft emission 
is the result of the contributions of different zones in the galactic bulge
with a large range of metallicities and densities (e.g. Baldi et al.~2006, 
Wang et al.~2009). Since we were interested in the hard component,
we froze the soft components to their best fit parameters in the subsequent analysis.\\
2. An intrinsic AGN emission consisting of an absorbed power law. The absorption
is modeled with a single photoelectric component, and then with a double component,
one of which partially covering the source (see below for details). \\
3. A cold reflection component, consisting of a continuum estimated with the XSPEC PEXRAV
model (Magdziarz \& Zdziarski~1995)  and an iron emission
line. \\

Since the spectral fits confirmed that no significant variation is present among the first
three almost consecutive observations, we repeated the analysis linking the parameters in these intervals. Hereinafter, we use the names OBS~1-3, OBS~4, and OBS~5 to refer to the 
three groups considered in the analysis.

\begin{table}
\caption{Results from the spectra analysis of {\em Chandra} data}
\centerline{\begin{tabular}{lcccccc}
Group  & $\Gamma$ & $N_{H}^a$ & R$^b$ & F$^c$ & $\chi^2$/d.o.f.\\
\hline
\multicolumn{6}{c}{Independent fits}\\
OBS 1-3         & 1.83$^{+0.47}_{-0.47}$ & 32$^{+7}_{-7}$       &1.7$^{+1.7}_{-0.8}$ & 2.2 & 176/215\\   
OBS 4           & 0.64$^{+0.78}_{-0.53}$ & 14.7$^{+2.3}_{-3.1}$ & $<$7               & 1.1 & 62/64\\
OBS 5           & 0.77$^{+0.66}_{-0.68}$ & 21$^{+12}_{-6}$      & $<$8               & 1.1 & 51/49\\ 
\hline 
\multicolumn{6}{c}{Constant reflection, 1.7$<$$\Gamma$$<$2.2}\\
OBS 1-3         & 1.7-2.2  & 29.6$^{+1.8}_{-1.7}$ &1.5$^{+1.0}_{-1.0}$& 1.8 & 306/326\\   
OBS 4           & 1.7-2.2  & 25.5$^{+3.3}_{-2.9}$ &2.0$^{+1.1}_{-1.0}$& 1.4 & 306/326 \\
OBS 5           & 1.7-2.2  & 37.6$^{+4.4}_{-4.6}$ &1.9$^{+2.0}_{-1.5}$& 1.4  & 306/326\\
\hline 
\end{tabular}}
$^a$ Column density in units of 10$^{23}$~cm$^{-2}$. 
$^b$: Ratio between the intrinsic continuum assumed in the reflection component and the
observed continuum. $^c$: unabsorbed 2-10~keV flux, in units of 10$^{-11}$~erg~cm$^{-2}$~s$^{-1}$.
\end{table}

The results from the spectral fitting are reported in Table~2~(top).
Even if the models
are acceptable from a statistical point of view ($\chi^2_\nu\sim1$ for all
the spectra), 
the present analysis 
is not satisfactory for two reasons:\\ 
1) the best fit continuum in the fourth and fifth
interval is extremely flat ($\Gamma$$\sim$0.7). Such a flat photon index has not been observed thus far
in any
non-reflection dominated AGN (see, e.g. the distribution of $\Gamma$ in nearby AGN
in Bianchi et al.~2009A, or Dadina et al.~2008), and is therefore unlikely to be real. 
Moreover, all the previous spectra of
UGC~4203 with high
statistics (Table~1), and the combined {\em Chandra} spectra of the first three interval, also 
having a much higher signal-to-noise, are well fitted with a fairly ``normal''
photon index, in the range 1.7-2.2. \\
2) The spectral parameters in OBS~4 and OBS~5 are poorly constrained, due to the 
high degeneracy between photon index, column density and reflection component. This is
evident from the large errors reported in Table~2. 
However, the $\Gamma$-N$_H$ contour plots 
(Fig.~2), show that a constant solution for all the intervals is ruled out at high statistical
significance. 

In order to overcome these limitations and better constrain the spectral parameters, we
repeated the spectral analysis fitting all the interval simultaneously, and requiring
a) a constant reflection component, and b) a continuum photon index in the range 1.7-2.2.
Not surprisingly (given the large contours in Fig.~2) the overall fit is completely
satisfactory ($\chi^2$=306/324~d.o.f.).
The results are listed in Table~2~(bottom) and shown in Fig.~3. 
Both variations in intrinsic flux and in absorbing column density are necessary to explain
the observed spectral variations. 
With these prescriptions, treating the three column densities as independent, the probability
of having no column density variation is P($\Delta$N$_H$=0)$<$10$^{-3}$). 
Specifically, 
the measured column densities in OBS~1-3 and OBS~4 are consistent, while the column density from
the fifth observation, N$_H$(OBS~5), is different from N$_H$(OBS~1-3) at a 99.5\% confidence level,
and from N$_H$(OBS~4) at a 99.98\% confidence level. The latter 
is the statistically most significant difference. The contour plot for N$_H$(OBS~4) and N$_H$(OBS~5)
are shown in Fig.~4.
  
Finally, repeating the fit assuming constant column density between OBS~1-3 and OBS~4,
we obtain N$_H$(OBS~1-4)=29.2$^{+1.5}_{-1.7}$$\times$10$^{23}$~cm$^{-2}$, 
and N$_H$(OBS~5)=37.7$_{-4.6}^{+5.2}$$\times$10$^{23}$~cm$^{-2}$.
The probability of no column density variation is obtained imposing N$_H$(OBS~5)=N$_H$(OBS~1-4).
 P($\Delta$N$_H$=0)=3$\times$10$^{-3}$.

We conclude that the column density variability between OBS~5 and the previous intervals is
highly significant. Changes in intrinsic flux are also present, 
and the continuum spectral slope may be also variable (within 
a physically acceptable range), but these components alone cannot reproduce the observations 
without a variable absorber.  


\begin{figure}
\includegraphics[width=6.2cm,angle=-90]{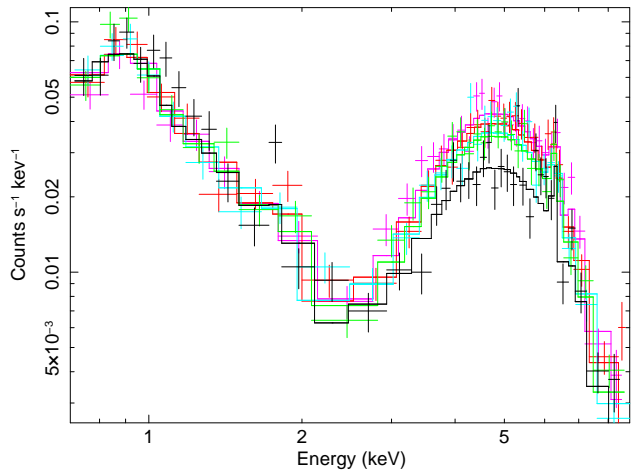}
\caption{{\em Chandra} 0.5-10~keV spectra of UGC~4203.
Note that the soft (E$<$2~keV) and hard (E$>$7‹keV) emission is constant within a few percent in all the
five observations, 
while significant variations occurred in the 2-6~keV range 
between the first four observations and the fifth one (in black). 
}
\end{figure}

\begin{figure}
\includegraphics[width=6.2cm,angle=-90]{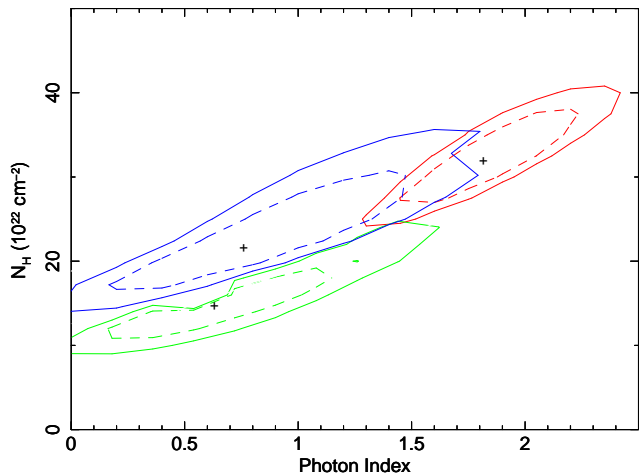}
\caption{Photon index-N$_H$ 1- and 2-$\sigma$ contours for OBS~1-3 (red), OBS~4 (green) and OBS~5 (blue). 
}
\end{figure}

\section{The structure of the X-ray absorber}
The results of our analysis show a column density variation in a time
scale $\Delta$T between 3 days (the time between the first and the fourth observation, during
which no variation has been found) and 17 days. This result presents a strong resemblance to
the column density variations found in the analogous {\em Chandra} campaign on NGC~1365 (Risaliti et al.~2007). Therefore, in the following we apply the same analysis as for NGC~1365.

\begin{figure}
\includegraphics[width=7.5cm,angle=-90]{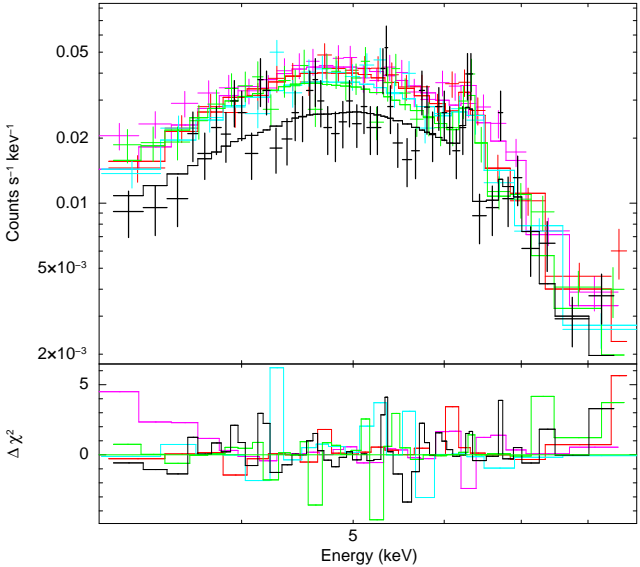}
\caption{3-10~keV spectrum, best fit model and $\chi^2$ residuals for the
five {\em Chandra} observations.}
\end{figure}

\begin{figure}
\includegraphics[width=8.5cm,angle=0]{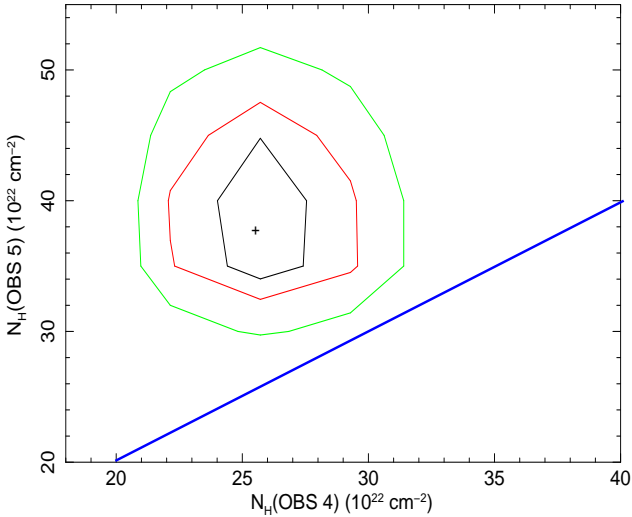}
\caption{Column density contours (1, 2~and 3~$\sigma$)  
from the best fit models to OBS~4 and OBS~5. The equal column density line is 
also plotted, for ease of comparison.  
}
\end{figure}

The variability  time scale can be easily translated into an estimate of the distance of
the absorbing cloud, assuming a rotation around the central black hole (BH) with
Keplerian velocity $V_K$. 
This estimate is subject to large uncertainties, connected to the values of the BH mass,
and of the cloud velocity, which are discussed below.\\
{\bf Black hole mass.} An estimate of the BH mass M$_{BH}$ can be
obtained from the known relations between M$_{BH}$ and spectral lines, bulge
luminosity, and bulge velocity dispersion. In the case of UGC~4203, discrepant
estimates are found in the literature: \\
1) M$_{BH}$-$\sigma$ relation. The relatively low bulge velocity
dispersion ($\sigma$=77~km~s$^{-1}$, Garcia-Rissmann et al.~2005) suggests 
log(M$_{BH}$/M$_\odot$)$\sim$6.5 (Bian \& Gu~2007). 
 The value obtained from the velocity dispersion imply a super-Eddington luminosity,
if a standard X-ray to bolometric correction is adopted, based on known
Quasar Spectral Energy Distributions (e.g. Risaliti \& Elvis~2004), and the
relation between optical to X-ray ratio and UV luminosity (Young et al.~2009).
We consider this scenario highly unlikely, mainly because of the rather constant
intrinsic X-ray flux (within a factor $\sim$2 of the one measured with {\em Chandra},
Table~1 and~2) over several years, which would imply a persistent super- or near-Eddington
luminosity for this long period.
On the other hand, the estimate of M$_{BH}$ based on velocity
dispersions may be strongly biased in this case by the presence of an unusually strong
starburst in the bulge of UGC~4203 ({Mazzalay 
\& Rodr{\'{\i}}guez-Ardila~2007).
\\
2) M$_{BH}$-bulge luminosity relation. We 
estimated M$_{BH}$ using the Marconi \& Hunt~(2003) relation between H-band bulge magnitudes
and BH masses, using the 2MASS H magnitude in the inner 4~kpc of the galaxy, and
the ratio between AGN and stellar contribution in H-band estimated by {Mazzalay 
\& Rodr{\'{\i}}guez-Ardila~(2007). We obtain
log(M$_{BH}$/M$_\odot$)$\sim$7.7.\\
\\
3) Virial mass estimates.
 The
reverberation mapping-calibrated relation of Vesteergard et al.~(2006)
suggests log(M$_{BH}$/M$_\odot$)$\sim$7.9 (Zhang et al.~2008). In this estimate
the width of the broad H$\beta$ line is measured in the polarized spectrum, while
the intrinsic luminosity at 5100~\AA~ is estimated from the [O III]~$\lambda$5007~\AA~
line.   
- The [O~III]-based luminosity is confirmed by an analogous estimate based on the
intrinsic X-ray luminosity, following the same prescriptions as mentioned above.
The measurement of FWHM(H$\beta$) is also quite solid, being based on a rather high quality
spectrum (Tran~1995). \\
  
Concluding, despite the large uncertainties in each of the methods summarized above,
we assume M$_{BH}$$\sim$5-7$\times$10$^7$~m$_\odot$ as the most likely mass estimate.

{\bf Cloud velocity.} the linear dimension of the X-ray source cannot be lower than a few gravitational
radii, $R_G$=GM$_{BH}$/c$^2$. Adopting $D_S$$\sim$5~R$_G$ as a lower limit, 
the velocity of the eclipsing cloud must therefore
satisfy the condition: V$_K$$\times$$\Delta$T$>$5~R$_G$.
Adopting the mass estimate discussed above, this translates into V$_K$$>$10$^3$~km~s$^{-1}$ $\Delta$T$_{10}$,
where $\Delta$T$_{10}$ is the time interval in units of 10~days. This value is clearly
in the range of the Broad Emission Line Clouds (BELCs) and well above that typical
of the standard, parsec-scale obscuring torus.\\
\subsection{Physical properties}
Based on the above estimates, and assuming Keplerian motion of the obscuring cloud,
we obtain a distance from the central black hole R$\sim$1-2$\times$10$^{17}$~cm.
The cloud linear size D$_C$ must be of the order of, or a few times larger, that that
of the X-ray source, i.e. D$_C$$\sim$1-5$\times$10$^{14}$~cm. The measured column density
$N_H$$\sim$10$^{23}$~cm$^{-2}$ imply a density of the order of 10$^9$~cm$^{-3}$.

We conclude that the X-ray absorber has all the typical physical parameters of a 
Broad Line Region cloud. Therefore, absorption in  X-rays and the broad optical/UV emission lines
are most likely due to the same physical component. We also note that UGC~4203 is optically classified
as a type~2 source, implying the presence of dust along the line of sight. This dust cannot be
associated to the Broad Line Region, which is well inside the dust sublimation radius, and must
therefore be located in a farther absorber. The nature of this second absorber may be a
galactic dust lane, and/or a "classical" parsec-scale torus. In either case, a contribution to
the observed X-ray absorption by this second component is expected. Within the limited statistics
of our {\em Chandra} observations it is not possible to disentangle these two components in the
spectral fits. The contribution to the observed column density may be as low as a few 
10$^{22}$~cm$^{-2}$, or as high as the lowest $N_H$ measured in past observations, i.e. 
$\sim$2$\times$10$^{23}$~cm$^{-2}$ (Table~1).

UGC~4203 is one more 
``Changing look'' AGN where the extreme X-ray variability
is proven to be due to absorption by BLR clouds. The same conclusion can be reached for a few
more sources where a similar variability has been found (NGC~4151, Puccetti et al.~2007, NGC~4388,
Elvis et al.~2004, NGC~7582, Bianchi et al.~2009B, Mrk~766, Risaliti et al.~2009).
On the other hand, it is likely that other bright sources with multiple X-ray observations
 do not show short-term absorption variability (even if no quantitative analysis has been 
done so far), and that other sources of X-ray absorption, quite far from the central source,
may be present (e.g., as discussed above, galactic dust lanes and/or parsec-scale tori). 
Therefore, 
since the physical link between BLR clouds and X-ray absorber has been now  
established for a good number
of sources, our next step will be the complete analysis of a representative sample, in order to
quantitatively estimate the relevance of short term variable X-ray absorption in local AGN.

\section*{Acknowledgements}
This research has made use of data obtained from the Chandra Data Archive and the Chandra Source Catalog, and software provided by the Chandra X-ray Center (CXC).
This work has been partially funded by NASA grants G08-9107X and G09-0105X.


\begin{thebibliography}{}
\bibitem[Arnaud(1996)]{1996ASPC..101...17A} Arnaud, K.~A.\ 1996,
Astronomical Data Analysis Software and Systems V, 101, 17
\bibitem[\protect\citeauthoryear{Baldi et al.}{2006}]{2006ApJS..162..113B} 
Baldi A., Raymond J.~C., Fabbiano G., Zezas A., Rots A.~H., Schweizer F., 
King A.~R., Ponman T.~J., 2006, ApJS, 162, 113
\bibitem[\protect\citeauthoryear{Bian 
\& Gu}{2007}]{2007ApJ...657..159B} Bian W., Gu Q., 2007, ApJ, 657, 159 
\bibitem[\protect\citeauthoryear{Bianchi et 
al.}{2009}]{2009A&A...495..421B} Bianchi S., Guainazzi M., Matt G., Fonseca Bonilla N., Ponti G., 2009A, A\&A, 495, 421 
\bibitem[\protect\citeauthoryear{Bianchi et 
al.}{2009}]{2009ApJ...695..781B} Bianchi S., Piconcelli E., Chiaberge M., 
Bail{\'o}n E.~J., Matt G., Fiore F., 2009B, ApJ, 695, 781 
\bibitem[Gilli et al.(2000)]{2000A&A...355..485G} Gilli, R., Maiolino, R.,
Marconi, A., Risaliti, G., Dadina, M., Weaver, K.~A., \& Colbert, E.~J.~M.\
2000, A\&A, 355, 485
\bibitem[Krolik \& Begelman(1988)]{1988ApJ...329..702K} Krolik, J.~H.~\&
Begelman, M.~C.\ 1988, ApJ, 329, 702
\bibitem[\protect\citeauthoryear{Dadina}{2008}]{2008A&A...485..417D} Dadina M., 2008, A\&A, 485, 417 
\bibitem[\protect\citeauthoryear{Elvis et al.}{2004}]{2004ApJ...615L..25E} 
Elvis M., Risaliti G., Nicastro F., Miller J.~M., Fiore F., Puccetti S., 
2004, ApJ, 615, L25
\bibitem[\protect\citeauthoryear{Garcia-Rissmann et 
al.}{2005}]{2005MNRAS.359..765G} Garcia-Rissmann A., Vega L.~R., Asari 
N.~V., Cid Fernandes R., Schmitt H., Gonz{\'a}lez Delgado R.~M., 
Storchi-Bergmann T., 2005, MNRAS, 359, 765
\bibitem[\protect\citeauthoryear{Guainazzi et 
al.}{2002}]{2002A&A...388..787G} Guainazzi M., Matt G., Fiore F., Perola G.~C., 2002, A\&A, 388, 787 
\bibitem[\protect\citeauthoryear{Guainazzi}{2002}]{2002MNRAS.329L..13G} 
Guainazzi M., 2002, MNRAS, 329, L13 
\bibitem[Magdziarz \& Zdziarski(1995)]{1995MNRAS.273..837M} Magdziarz,
P.~\& Zdziarski, A.~A.\ 1995, MNRAS, 273, 837
\bibitem[\protect\citeauthoryear{Marconi 
\& Hunt}{2003}]{2003ApJ...589L..21M} Marconi A., Hunt L.~K., 2003, ApJ, 589, L21 
\bibitem[Matt, Guainazzi, \& Maiolino(2003)]{2003MNRAS.342..422M} Matt, G.,
Guainazzi, M., \& Maiolino, R.\ 2003, MNRAS, 342, 422
\bibitem[\protect\citeauthoryear{Matt et 
al.}{2009}]{2009A&A...496..653M} Matt G., Bianchi S., Awaki H., Comastri A., Guainazzi M., Iwasawa K., Jimenez-Bailon E., Nicastro F., 2009, A\&A, 496, 653 
\bibitem[\protect\citeauthoryear{Mazzalay 
\& Rodr{\'{\i}}guez-Ardila}{2007}]{2007A&A...463..445M} Mazzalay X., Rodr{\'{\i}}guez-Ardila A., 2007, A\&A, 463, 445
\bibitem[\protect\citeauthoryear{Piconcelli et 
al.}{2007}]{2007A&A...466..855P} Piconcelli E., Bianchi S., Guainazzi M., Fiore F., Chiaberge M., 2007, A\&A, 466, 855 
\bibitem[Puccetti et al.(2007)]{2007MNRAS.377..607P} Puccetti, S., Fiore,
F., Risaliti, G., Capalbi, M., Elvis, M.,
\& Nicastro, F.\ 2007, MNRAS, 377, 607
\bibitem[Risaliti, Elvis, \& Nicastro(2002)]{2002ApJ...571..234R} Risaliti,
G., Elvis, M., \& Nicastro, F.\ 2002, ApJ, 571, 234
\bibitem[\protect\citeauthoryear{Risaliti 
\& Elvis}{2004}]{2004ASSL..308..187R} Risaliti G., Elvis M., 2004, ASSL, 308, 187 
\bibitem[Risaliti et al.(2005)]{2005ApJ...623L..93R} Risaliti, G., Elvis,
M., Fabbiano, G., Baldi, A., \& Zezas, A.\ 2005a, ApJ, 623, L93 (R05A)
\bibitem[Risaliti et al.(2007)]{2007ApJ...659L.111R} Risaliti, G., Elvis,
M., Fabbiano, G., Baldi, A., Zezas, A.,
\& Salvati, M.\ 2007, ApJ, 659, L111
\bibitem[\protect\citeauthoryear{Risaliti et 
al.}{2009}]{2009MNRAS.393L...1R} Risaliti G., et al., 2009A, MNRAS, 393, L1 
\bibitem[\protect\citeauthoryear{Risaliti et 
al.}{2009}]{2009ApJ...696..160R} Risaliti G., et al., 2009B, ApJ, 696, 160 
\bibitem[\protect\citeauthoryear{Young, Elvis, 
\& Risaliti}{2009}]{2009arXiv0911.0474Y} Young M., Elvis M., Risaliti G., 2009, arXiv, arXiv:0911.0474 
\bibitem[\protect\citeauthoryear{Zhang, Bian, 
\& Huang}{2008}]{2008A&A...488..113Z} Zhang S.-Y., Bian W.-H., Huang K.-L., 2008, A\&A, 488, 113 
\bibitem[\protect\citeauthoryear{Wang et al.}{2009}]{2009ApJ...694..718W} 
Wang J., Fabbiano G., Elvis M., Risaliti G., Mazzarella J.~M., Howell 
J.~H., Lord S., 2009, ApJ, 694, 718
\end{thebibliography}
\end{document}